\documentclass{aa}
\usepackage{psfig}
\def\la{\;
\raise0.3ex\hbox{$<$\kern-0.75em\raise-1.1ex\hbox{$\sim$}}\; }
\def\ga{\;
\raise0.3ex\hbox{$>$\kern-0.75em\raise-1.1ex\hbox{$\sim$}}\; }
\begin{document}
\thesaurus{12(09.13.2; 02.12.3; 11.17.1; 11.17.4 Q0000--2620)}
\title{
UVES observations of QSO 0000--2620~: Molecular
hydrogen abundance in the damped Ly$\alpha$ system at
$z_{\rm abs} = 3.3901$\thanks{ Based on public data released from
the UVES commissioning at the VLT
Kueyen telescope, European Southern Observatory, Paranal, Chile. } 
}
\author{S. A. Levshakov\inst{1}
\and P. Molaro\inst{2}
\and M. Centuri\'on\inst{2}
\and S. D'Odorico\inst{3}
\and P. Bonifacio\inst{2}
\and G. Vladilo\inst{2}}
\offprints{S. A. Levshakov}
\mail{lev@astro.ioffe.rssi.ru}
\institute{
Department of Theoretical Astrophysics,
Ioffe Physico-Technical Institute, 
194021 St. Petersburg, Russia
\and
Osservatorio Astronomico di Trieste, Via G.B. Tiepolo 11, 34131, 
Trieste, Italy
\and
European Southern Observatory, 85748 Garching bei M\"unchen, Germany}
\date{Received 24 May 2000 / Accepted 00 July  2000}
\titlerunning{H$_2$ at $z_{\rm abs} = 3.3901$ toward Q~0000--2620}
\authorrunning{S.A. Levshakov et al.}
\maketitle
\begin{abstract}
We have discovered molecular 
hydrogen in a fourth quasar damped Ly$\alpha$ system
(hereafter DLA).
The UVES spectrograph on the 8.2m ESO Kueyen telescope
has allowed the detection of H$_2$ in gas with low metallicity,
$Z/Z_\odot \simeq 10^{-2}$, 
and high neutral hydrogen column density,
$N(\ion{H}{i}) \simeq 2.6\times10^{21}$ cm$^{-2}$,
at redshift $z_{\rm abs} = 3.3901$ toward QSO 0000--2620.
The measured H$_2$ fractional abundance of 
$f({\rm H}_2) \simeq 4\times10^{-8}$
is lower than a typical value for Galactic interstellar clouds of high
$N(\ion{H}{i})$ column density by a factor of $(2-3)\times10^6$.
Since H$_2$ molecules are formed efficiently on dust grains,
it implies that the dust condensation in this DLA is negligible,
and hence the abundances derived 
from metal absorption lines are the actual ones.
The obtained $f({\rm H}_2)$ value leads to an estimate of the
dust number density of
$\langle n_{\rm d} \rangle_{\rm DLA} \sim 10^{-3}\,
\langle n_{\rm d} \rangle_{\rm ISM}$,
which is consistent with the dust-to-gas ratio 
$\tilde{k} \leq 1.6\times10^{-3}$ derived independently from the
[Cr/Zn] and [Fe/Zn] ratios.

\end{abstract}

\section{Introduction}

In this work we continue our chemical composition analysis 
of the $z_{\rm abs} = 3.3901$ damped Ly$\alpha$ system 
from the spectrum of the bright quasar
Q0000--2620 (V = 17.5, $z_{\rm em} = 4.108$) observed with the
UVES spectrograph (Dekker et al. 2000) at the 8.2m ESO Kueyen
telescope during the first commissioning in October 1999
(Molaro et al. 2000, hereafter Paper~I). 

Paper I was mainly concerned with precise measurements of the
oxygen and zinc abundances. The obtained O and Zn metallicities
together with other $\alpha$-chain and iron-peak element abundances
present a pattern of a galactic chemical composition different
from that of the Milky Way. In particular, it was found that 
the relative abundances of $\alpha$-chain and iron-peak elements,
[O,Si,S/Cr,Fe,Zn], are significantly lower than analogous ratios
in Galactic stars with comparable metallicities. 
The measured Zn abundance,
[Zn/H] = $-2.07 \pm 0.1$ dex\footnote{Using the customary
definition [X/H] = log (X/H) -- log (X/H)$_\odot$},
is the lowest among DLAs, which implies that the $z_{\rm abs} = 3.3901$
intervening galaxy is in the early stages of its chemical evolution.
This suggestion is consistent with the similar abundances found for the 
volatile (Zn) and refractory (Cr and Fe) elements,
[Cr/H] = $-1.99 \pm 0.1$ dex and [Fe/H] = $-2.04 \pm 0.1$ dex,
which means, in turn, that the Cr and Fe atoms may not be locked up in
dust grains and that the dust-to-gas ratio in this DLA may be much lower
if compared with Galactic diffuse clouds with the same neutral
hydrogen column densities, $N(\ion{H}{i}) \simeq 2.5\times10^{21}$
cm$^{-2}$.

The low dust-to-gas ratio in this system suggested in Paper~I
is consistent with the absence
of significant H$_2$ absorption at $z_{\rm abs} = 3.3901$ in the Lyman 
${\rm B}^1 \Sigma^{+}_{\rm u} \leftarrow {\rm X}^1 \Sigma^{+}_{\rm g}$
and Werner 
${\rm C}^1 \Pi^{\pm}_{\rm u} \leftarrow {\rm X}^1 \Sigma^{+}_{\rm g}$
bands (Levshakov et al. 1992, hereafter LCFB)
as well as with the absence of CO lines from the fourth positive band
${\rm A}^1 \Pi \leftarrow {\rm X}^1 \Sigma^{+}$ (Lu et al. 1999).

It should be noted that in the Galactic interstellar medium, 
sightlines with $N(\ion{H}{i}) \ga 10^{21}$ cm$^{-2}$
show high fractional abundances of hydrogen in molecular form,
$f({\rm H}_2) \sim 0.1$ (Savage et al. 1977), where
$f({\rm H}_2) = 2N({\rm H}_2)/N({\rm H})$ with $N({\rm H}_2)$ and  
$N({\rm H})$ being the total column densities of H$_2$ and H,
respectively.
Such molecular clouds with $N({\rm H}_2) \ga 10^{19}$ cm$^{-2}$
provide an effective self-shielding in the Lyman and Werner bands
preventing UV photons from penetrating the interior of the clouds
and thus lowering photodissociation rates of H$_2$ and CO molecules.

In contrast with the ISM diffuse clouds, molecular hydrogen in DLAs
shows different behavior. Namely, we do not observe a pronounced
transition from low to high molecular fractions at total hydrogen
column density near $5\times10^{20}$ cm$^{-2}$ as found in the 
Milky Way disk by Savage et al. (1977).
The first H$_2$ system identified at
$z_{\rm abs} = 2.8112$ toward Q0528--250 by Levshakov \& Varshalovich
(1985) gives very low amount of H$_2$, 
$f({\rm H}_2) \simeq 5.4\times10^{-5}$, while 
$N(\ion{H}{i}) \simeq 2.2\times10^{21}$ cm$^{-2}$
(see also Srianand \& Petitjean 1998 and references cited therein). 
However, the second and the third H$_2$ systems found at
$z_{\rm abs} = 1.97$ toward Q0013--004 (Ge \& Bechtold 1997) and
at $z_{\rm abs} = 2.34$ toward Q1232+0815 (Ge et al. 2000)
show high H$_2$ abundances, $f({\rm H}_2) \simeq 0.22$ and 0.07,
respectively (the corresponding neutral hydrogen column densities
are equal to $6.4\times10^{20}$ cm$^{-2}$ and 
$8.0\times10^{20}$ cm$^{-2}$).
Other DLAs have led so far only to upper limits,
$f({\rm H}_2) \la 2\times10^{-4}$, with the lowest bound of
$9\times10^{-7}$ at $z_{\rm abs} = 2.47$ toward Q1223+17 (Bechtold 1999).  

LCFB derived an upper limit of $f({\rm H}_2) < 3\times10^{-6}$
for the DLA at $z_{\rm abs} = 3.3901$ toward Q0000--2620.
The correlation between the H$_2$ fractional abundance and the
[Cr/Zn] ratio 
found by Ge \& Bechtold (1999; see their Fig.~1) suggests
that $f({\rm H}_2) \la 10^{-7}$ if
[Cr/Zn] $\sim 0$, as measured in Paper~I.
This motivated us to search for the H$_2$ transitions
in the UVES spectrum already discussed in Paper~I.
We show below that the S/N of this spectrum is indeed
adequate to measure H$_2$, thus making this absorber the
fourth one for which such a measure is available.

\section{Data analysis and Results }

\begin{figure}
\hspace{0.3cm}\psfig{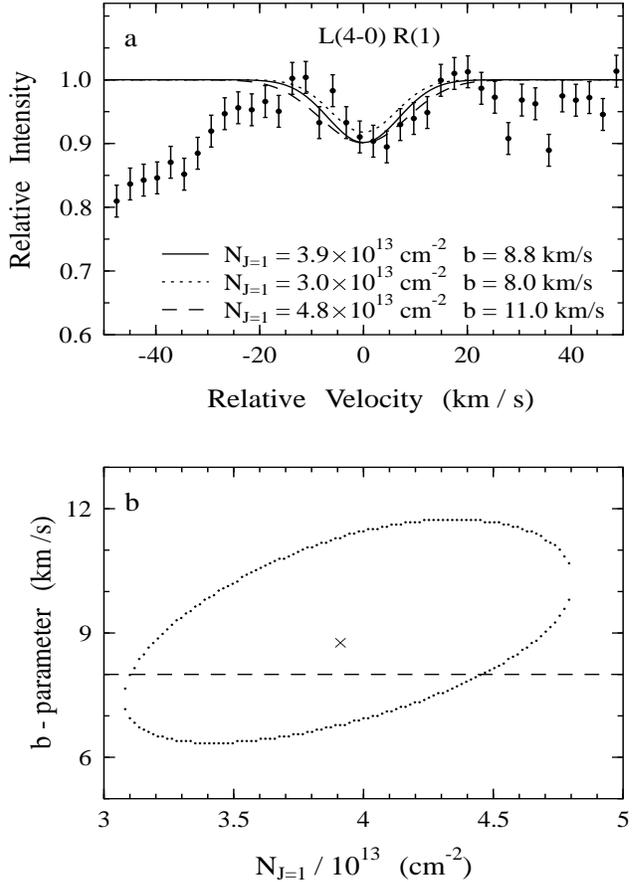}
\vspace{-2.0cm}
\caption[]{({\bf a}) -- Observed normalized intensities 
(dots with error bars) vs radial velocities given
relative to  $z_{\rm abs} = 3.390127$. 
The rest frame wavelength of the L(4-0)R(1) line and its oscillator strength
(1049.958 \AA\, and 0.0160, respectively) 
were taken from Morton \& Dinerstein (1976).
Smooth lines are the synthetic
spectra obtained from the $\chi^2$ minimization as described in the text.
({\bf b}) -- Confidence region in the `$N_{J=1}({\rm H}_2)-b$' plane.
The contour above the dashed horizontal line represents 70\% accepted confidence
level. The horizontal line restricts $b$-values which are less than
the lowest $1\sigma$ boundary for the $b$-value of the heaviest element
from the $z_{\rm abs} = 3.3901$ system,
$b_{\ion{Zn}{ii}} \geq 8.0$ km~s$^{-1}$ (Paper~I). The cross marks
the point of maximum likelihood for the best fit shown by the solid line
in panel {\bf a}
}
\end{figure}

\subsection{ Observations and the H$_2$ identification }

Spectroscopic observations of Q0000--2620 are described in detail in Paper~I.
The spectrum was obtained with the rms uncertainty
of the wavelength calibration $\delta \lambda \leq 0.6$ km~s$^{-1}$,
the velocity resolution of FWHM $\simeq 6$ km~s$^{-1}$ 
(the corresponding bin size equals 2.4 km~s$^{-1}$) 
and the signal-to-noise
ratio of S/N $\simeq 40$ (per pixel)  
in the range $\lambda\lambda = 4605 - 4615$ \AA\,
which allows us to detect the L(4-0)R(1) line at the
expected position, $\lambda_{\rm obs} = 4609.4$ \AA.\,
The absence of this line in the 1~\AA\, resolution
spectrum obtained with the Multiple Mirror Telescope (MMT)
in 1988 (LCFB) has led to the limit 
$W_{\rm rest} < 114$ m\AA\, $(3\sigma)$ for the equivalent width of
the  L(4-0)R(1) line. The UVES spectrum shown in Fig.~1a
provides $W_{\rm rest} \simeq 6$ m\AA.\, The local continuum level
in the vicinity of the L(4-0)R(1) line was additionally controlled by
the optical depth value of the Lyman-limit discontinuity
$(\tau_{\rm c} = 0.003)$ from the metal system at
$z_{\rm abs} = 4.061$ (Savaglio et al. 1996).
This is illustrated in Fig.~2 where the
estimated continuum level between 4555 \AA\, and 4660 \AA\, is shown
by the dotted line. The vertical line at $\lambda = 4614.39$ \AA\,
marks the Lyman-limit position for the $z_{\rm abs} = 4.061$ system. 
The expected optical depth in the continuum at $\lambda = 4609$~\AA\,
is equal to
$\tau_{4609} = \tau_{\rm c}\,(4609/4614)^3 \simeq 0.003$.

\begin{figure*}
\psfig{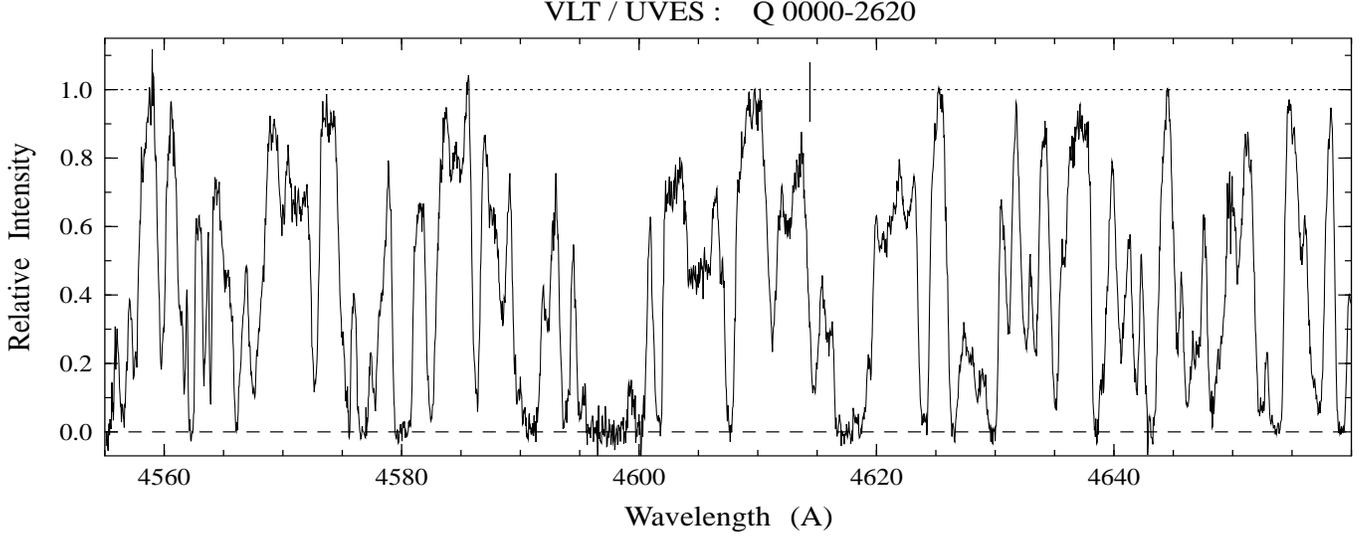}
\vspace{0.5cm}
\caption[]{
A portion of the Q~0000--2620 spectrum in the vicinity of the
H$_2$ L(4-0)R(1) line ($\lambda_{\rm obs} = 4609.4$ \AA). 
The Lyman-limit discontinuity of the metal system at $z_{\rm abs} = 4.061$
($\tau_{\rm c} = 0.003$) is marked by the vertical line
(see text for details)
}
\label{f:2}
\end{figure*}

We have thoroughly investigated all possible 
coincidences of the L(4-0)R(1) position with metal
absorptions from other systems observed toward Q0000--2620.
Only one such case was found. The position of the
\ion{C}{ii}$\lambda 903.6235$ ($f_{\rm abs} = 0.168$)
line from the metal system
at $z_{\rm abs} = 4.10106$ detected by Savaglio et al. (1997)
differs from the center of the L(4-0)R(1) line by only 
$-0.7$ km~s$^{-1}$. 
If the feature observed ($W_{\rm obs} = 26$ m\AA) were due to
this \ion{C}{ii} absorption, we would obtain $W_{\rm rest} = 5$ m\AA\,
and $N(\ion{C}{ii}) = 4.27\times10^{12}$ cm$^{-2}$ 
using the linear part
of the curve of growth. This would imply an abundance
[C/H] $\simeq + 1.68$ which is unlikely, due to the high redshift
of the absorber. We remark that even assuming a solar carbon
abundance, [C/H] = 0, and the unrealistic suggestion that all
carbon is in the \ion{C}{ii} form (Savaglio et al. have
detected \ion{C}{iv} in this system which is not a DLA)
we find that $W_{\rm obs}(\ion{C}{ii}) = 0.5$ m\AA.\,
This result implies that the \ion{C}{ii} absorption
at $z_{\rm abs} = 4.10106$ does not affect significantly
the H$_2$ profile.

The identification of other H$_2$ transitions from the Lyman and Werner
bands is hampered by blending with numerous Ly$\alpha$ forest lines.
Therefore, to obtain quantitative estimates of the total molecular
hydrogen column density and the excitation temperature, $T_{\rm ex}$,
we choose small continuum `windows' which are less obscured by the
Ly$\alpha$ forest absorption features and where some H$_2$ lines
are expected. The following is a description of the adopted procedure.

\subsection{ Column densities and the excitation temperature } 

The inferred H$_2$ column density, $N({\rm H}_2) = \sum_J N_J$,
depends on the populations of the first rotational levels
$(J = 0, 1, 2, ...)$ in the ground state, $T_{\rm ex}$, and the
Doppler $b$-parameter. In thermal equilibrium, the ratio of the
column densities $N_J$ is given by
\begin{equation}
\frac{N_J}{N_0} = \frac{g_J}{g_0} \exp \left[
- \frac{B_{\it v}J(J+1)}{T_{\rm ex}} \right]\; ,
\label{eq:E1}
\end{equation} 
where the statistical weight of a level $J$ is $3(2J+1)$ for odd $J$
and $(2J+1)$ for even $J$ (the ortho- and para-H$_2$, respectively),
and the constant $B_{\it v} = 85.36$~K for the vibrational ground state.

The central optical depth, $\tau_0$, for the resolved absorption
line is determined by (e.g., Spitzer 1978)
\begin{equation}
\tau_0 = 1.497\times10^{-15} \lambda_0 f_{\rm abs} 
\frac{N}{b}\; ,
\label{eq:E2}
\end{equation} 
where $f_{\rm abs}$ is the oscillator strength of the transition,
$\lambda_0$ its wavelength in \AA, $N$ the total column density
in cm$^{-2}$, and $b$ the Doppler parameter in km~s$^{-1}$.
For the H$_2$ lines, wavelengths and oscillator strengths were taken
from the list of Morton \& Dinerstein (1976).

To estimate the ortho-H$_2$ column density, $N_1$, and the $b$-parameter,
we fit theoretical Voigt profiles to the observed intensities via
$\chi^2$ minimization using a fixed value of $z_{\rm abs} = 3.390127$
which corresponds to the redshift of the isolated 
\ion{Ni}{ii}$\lambda 1709$
absorption line (Paper~I).
During the fitting procedure the theoretical profiles are convolved
with a Gaussian of the instrumental width of 6 km~s$^{-1}$.
We tried to find a satisfactory result with the reduced $\chi^2$ per
degree of freedom of $\chi^2_{\rm min} \leq \chi^2_{\nu,0.30}$
(where $\chi^2_{\nu,0.30}$ is the expected $\chi^2$ value for $\nu$
degrees of freedom at the credible probability of 70 \%).
We used the velocity range $| \Delta v | \leq 20$ km~s$^{-1}$
where we have $\chi^2_{14,0.30} = 1.159$.
The best fit with $\chi^2_{\rm min} = 0.80$ is shown in Fig.~1a
by the solid curve, whereas points with error bars give the
normalized intensities (the corresponding S/N $\simeq 40$).

\begin{figure}
\hspace{0.3cm}\psfig{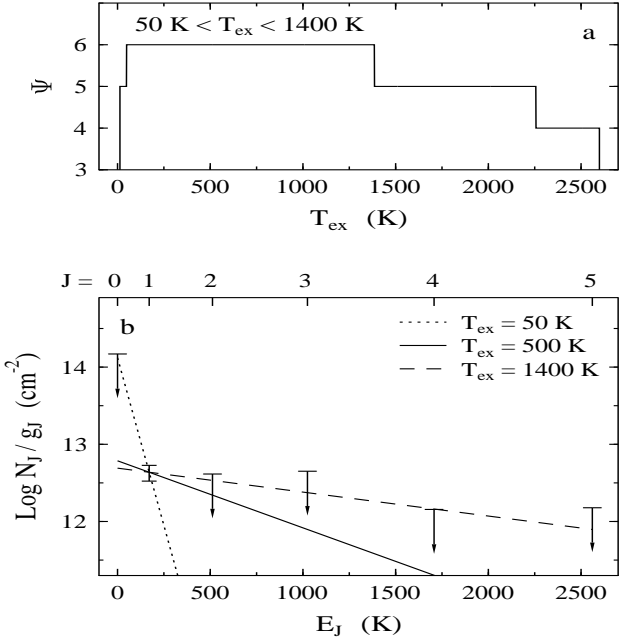}
\vspace{-3.2cm}
\caption[]{({\bf a}) -- $\Psi$- function defined in Eq. 
(\ref{eq:E4}) vs the excitation temperature  $T_{\rm ex}$. 
The values of $\Psi = 6$ restrict the range for acceptable  $T_{\rm ex}$ 
as illustrated in panel {\bf b}.
({\bf b}) -- Population density in H$_2$ rotational levels observed
in the $z_{\rm abs} = 3.3901$ system. The column density $N_{J=1}$
for H$_2$ molecules in level $J=1$ per cm$^2$ (or the upper limits
of $N_J$ for $J = 0, 2 - 5$), divided by the statistical weight,
$g_{J}$, is plotted on a logarithmic 
scale against the excitation energy, $E_J$,  
in K. The corresponding values of $J$ are shown in the horizontal scale at the
top. The points have been fitted with one straight line using Eq.(\ref{eq:E4}),
and a value of $T_{\rm ex}$, obtained from the usual Boltzmann formula. 
The shown straight lines correspond to $T_{\rm ex} = 50, 500$, and 1400~K
}
\end{figure}

The best solution yields $N_1 = 3.9\times10^{13}$ cm$^{-2}$
and $b = 8.8$ km~s$^{-1}$. Of course, our analysis gives a certain
range for allowable $N_1$ and $b$. 
This is shown in Fig.~1b, where the 70~\%
confidence region for this pair of physical parameters is restricted
by the upper part of the ellipse over the line $b = 8$ km~s$^{-1}$.
The second part of the ellipse under this horizontal dashed line
may be excluded from consideration 
since the $b$-value of the heaviest element \ion{Zn}{ii}
equals $9.3 \pm 1.3$ km~s$^{-1}$ (Paper~I), otherwise one would have
a zone where $b_{{\rm H}_2} < b_{\ion{Zn}{ii}}$.   
The cross in Fig.~1b marks the point of maximum likelihood for the
best fit.
Two additional curves in Fig.~1a (dotted and dashed) show the
L(4-0)R(1) profiles for the limiting values of $N_1 = 3.0\times10^{13}$
cm$^{-2}$ and $4.8\times10^{13}$ cm$^{-2}$, respectively. Both of them
have $\chi^2_{\rm min} = 1.06$. To summarize, we set
$N_1 = (3.9 \pm 0.9)\times10^{13}$ cm$^{-2}$ and 
$b_{{\rm H}_2} = 8.8^{+3.0}_{-0.8}$ km~s$^{-1}$.

Let us now turn to the estimation of the total $N({\rm H}_2)$.
As mentioned above the total molecular hydrogen column density
depends on $T_{\rm ex}$. Typical ISM diffuse molecular clouds
have a mean $(J=0)/(J=1)$ excitation temperature of
$\langle T_{01} \rangle = 77 \pm 17$ K, with the bulk lying
in the range from 45 to 128 K (Savage et al. 1977).
The excitation temperature $T_{01}$ reflects the relative
populations of $J=0$ and 1 rotational levels which are controlled
by thermal particle collisions. Therefore $T_{01}$
is approximately equal to the kinetic temperature, $T_{\rm kin}$,
of the cloud. The population of the higher rotational levels 
are established by collisions, UV pumping and radiative cascades
after photo-absorption to the Lyman and Werner bands
and thus $T_{\rm ex}$ for the higher levels may differ from
$T_{01}$ in dense molecular clouds (see, e.g. a review by Dalgarno 1976).
However, for tenuous clouds, a unique temperature for all the levels 
can be found suggesting that the rotational population could
be dominated by collisions in a gas whose kinetic temperature
is near $T_{\rm ex}$. For example, the excitation temperature
of $1120 \pm 80$ K toward $\zeta$~Pup was measured 
for $J = 0 - 7$ by Morton \& Dinerstein (1976). 

There is also a tendency toward low 
excitation temperatures $T_{\rm ex} \sim 100$ K for the clouds
with $f({\rm H}_2) > 0.05$, whereas clouds with  $f({\rm H}_2) < 0.05$
tend to higher $T_{\rm ex}$ (Spitzer \& Jenkins 1975).
Note that spin temperatures $\ga 10^3$ K
have been measured in
DLAs with $N(\ion{H}{i}) > 10^{21}$ cm$^{-2}$ 
for which 21~cm line measurements are available
(de Bruyn et al. 1996, Carilli et al. 1996). 
These measurements assume, however, a similar covering
factor of the optical and radio background source. 

The excitation temperature can be estimated directly from the
comparison of the population density in H$_2$ rotational levels
or indirectly from the thermal width of the H$_2$ lines assuming
that $T_{\rm ex}$ is approximately equal to the kinetic
temperature of the cloud. 
For the DLA in question, we know exactly that the line widths do
not reflect the thermal motion since all species from the lightest
H$_2$ to the heaviest Zn show approximately the same $b$-values
(see Paper~I). We also have not accurate measurements of $N_J$
for different $J$. Therefore, to estimate $T_{\rm ex}$ we applied
another method.

\begin{figure}
\hspace{0.0cm}\psfig{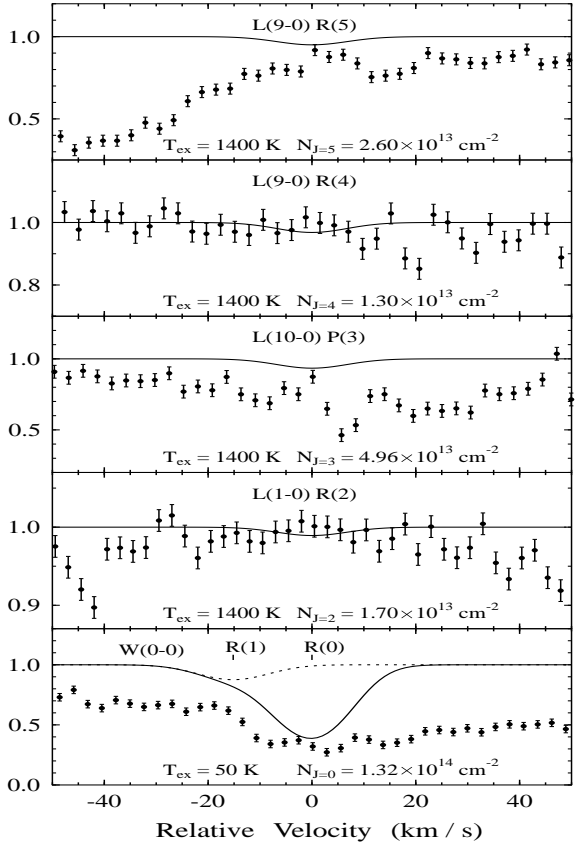}
\vspace{0.5cm}
\caption[]{Velocity plots of the UVES normalized data of Q0000--2620 
(dots and 1$\sigma$ error bars) obtained by Molaro et al. (2000) 
and the synthetic H$_2$ spectra convolved with the instrumental
resolution of FWHM = 6 km~s$^{-1}$ (solid curves), corresponding
to the limits with $T_{\rm ex} = 50$~K for $J = 0$ (bottom panel)
and $T_{\rm ex} = 1400$~K for $J = 2-5$. The dotted curve in 
the bottom panel shows the W(0-0)R(1) profile for 
$N_{J=1} = 3.9\times10^{13}$ cm$^{-2}$; the Doppler
parameter equals $b_{{\rm H}_2} = 8.8$ km~s$^{-1}$ for all synthetic
spectra (see also Fig.~1a)
}
\end{figure}

We choose the continuum `windows' which are sensitive to the presence
of the H$_2$ absorption and set upper limits for $N_J$ using
the inequality
\begin{equation}
\tau^{{\rm H}_2}_0 < \tau^{\rm obs}_0\; ,
\label{eq:E3}
\end{equation} 
with $\tau^{\rm obs}_0$ being the observed limit at the expected
H$_2$ transition and $\tau^{{\rm H}_2}_0$ the quantity calculated
from (2). 
In this estimations we used 
$b_{{\rm H}_2} = 8.8$ km~s$^{-1}$.
The obtained limits are depicted by short horizontal bars in Fig.~3b.

Now we know the upper limits $N^{\rm lim}_J$ for $J = 0,2,3,4,5$ and
the measured $N_1$ value. To estimate $T_{\rm ex}$ consider
the following function~: 
\begin{equation}
\Psi (T_{\rm ex}) = \sum_J \psi_J\; ,
\label{eq:E4}
\end{equation} 
where for $J = 0,2,3,4,5$ we have
\[ \psi_J = \left\{
\begin{array}{rl}
1, & \mbox{if } N^{\rm calc}_J < N^{\rm lim}_J \\
0, & \mbox{otherwise  ,}
\end{array} \right. \]
and for $J = 1$
\[ \psi_J = \left\{
\begin{array}{rl}
1, & \mbox{if } |N^{\rm calc}_1 - N_1| \leq \delta N \\
0, & \mbox{otherwise ,}
\end{array} \right. \]
where $\delta N = 0.9\times10^{13}$ cm$^{-2}$.

\smallskip\noindent
Here $N^{\rm calc}_J$ is calculated as
\begin{equation}
N^{\rm calc}_J = N_1 \frac{g_J}{g_1}
\exp\left[ - \frac{B_{\it v}J(J+1)}{T_{\rm ex}} +
\frac{2B_{\it v}}{T_{\rm ex}} \right]\; .
\label{eq:E5}
\end{equation} 

It follows from the definition of $\Psi(T_{\rm ex})$ that
the allowable range for $T_{\rm ex}$ should correspond to the
maximum value of $\Psi$ (here equals 6).
The shape of $\Psi$ as function of $T_{\rm ex}$ is shown in Fig.~3a
which readily gives the range 50~K $< T_{\rm ex} <$ 1400~K.
The range is rather wide since we have only one unblended H$_2$ line.
It should be noted, however, that the obtained limits do not 
contradict the above mentioned observational estimations of $T_{\rm ex}$.

Fig.~3b gives additional illustrations for our procedure. 
Here we show the measured $(J = 1)$ and estimated $(J \neq 1)$
column densities against energies of the low rotational levels.
Three different lines represent solutions for $T_{\rm ex} = 50$, 500
and 1400~K. As seen, at low $T_{\rm ex}$  the most sensitive are
the $J = 0$ and $J = 1$ transitions, whereas at high $T_{\rm ex}$
the most stringent restriction is set by the 
populations of the $J = 2$ and $J = 4$ rotational levels.

These two regimes are shown in Fig.~4 where dots with error bars
are the observed normalized intensities within the chosen windows,
and the solid curves show the H$_2$ synthetic profiles calculated
with $b_{{\rm H}_2} = 8.8$ km~s$^{-1}$ and the limiting column
densities listed in the corresponding panels. 

If $T_{\rm ex}$ lies between 50~K and 1400~K, the total molecular
hydrogen column density $N({\rm H}_2)$ ranges from $5.4\times10^{13}$
cm$^{-2}$ to $1.7\times10^{14}$ cm$^{-2}$ (see Fig.~5) where the lower
value is obtained at $T_{\rm ex} = 2 B_{\it v}$.
Thus we may set $\log N({\rm H}_2) = 13.98 \pm 0.25$.
With $\log N(\ion{H}{i}) = 21.41 \pm 0.08$\, taken from Lu et al.
(1996), and assuming that the molecular gas coincide with the 
\ion{H}{i} (i.e. filling factor is 1),
one finds 
$\log f({\rm H}_2) = -7.43 \pm 0.26$ or
$f({\rm H}_2) \simeq 4\times10^{-8}$.

Since H$_2$ molecules are formed more efficiently on dust grains
in high \ion{H}{i} column density clouds,
we can conclude that the dust number density in the 
$z_{\rm abs} = 3.3901$ absorbing material is 
lower than in the ISM clouds with
$N(\ion{H}{i}) \ga 10^{21}$ cm$^{-2}$,
provided the dust destruction rate is the same.

\subsection{ Dust Content }

The presence of dust in DLAs is illustrated in Fig.~6 where
the relation between $\log f({\rm H}_2)$ and  relative heavy element
depletion, [Cr/Zn], is shown. 
In DLAs, the amount of dust is usually estimated from the ratio
$N(\ion{Cr}{ii})/N(\ion{Zn}{ii})$ assuming that Zn
is undepleted (e.g. Pettini et al. 1994). 

The measured data for the four extragalactic molecular clouds 
(dots in Fig.~6)
can be very well fitted with a linear law~:
$\log f({\rm H}_2) \simeq -5.5\, [{\rm Cr}/{\rm Zn}] - 7.1$,\, 
which means that these quantities are strongly correlated
(dashed line in Fig.~6).
Another four DLAs with high neutral hydrogen column densities
where the H$_2$ absorption has not yet been detected (see Table~1)
are marked by vertical arrows. Crosses in Fig.~6 are 
the Galactic interstellar medium measurements taken from
the compilation of Roth \& Blades (1995).
By the horizontal lines we mark the H$_2$ measurements in the
Magellanic Clouds (Richter 2000). 
Unfortunately, the corresponding [Cr/Zn] data 
are not available yet, 
therefore, to restrict their plausible range
we took the limiting values of [Cr/Zn] $= -1.09$ and $-0.72$
obtained
from different sightline measurements in the Magellanic Clouds 
by Roth \& Blades (1997). 

\begin{figure}
\hspace{0.3cm}\psfig{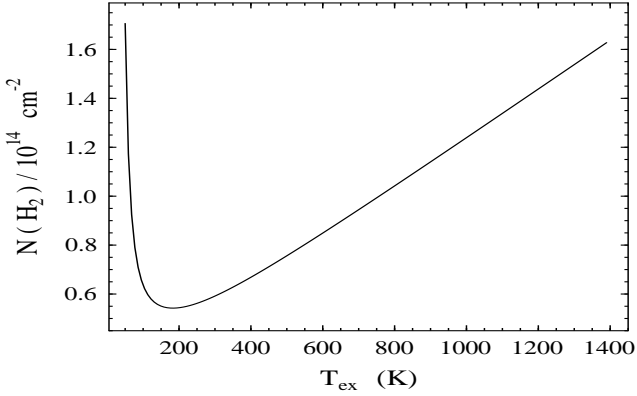}
\vspace{0.3cm}
\caption[]{ The total molecular hydrogen column density $N({\rm H}_2)$
vs $T_{\rm ex}$ for a fixed value $N_{J=1} = 3.9\times10^{13}$ cm$^{-2}$ 
}
\end{figure}

Qualitatively, the DLA data, including the 
$f({\rm H}_2)$ upper limits, exhibit the same tendency as the
Galactic ISM diffuse clouds, i.e. clouds with large amount of H$_2$
show higher heavy element depletion. 
The linear correlation among the DLA systems is then a kind
of upper envelope of all the different types of data points.
The fact that the ISM measurements are found at higher Cr
depletion than the DLA data is probably due to the higher
metallicity and dust content of interstellar clouds.
In fact, there is evidence for a correlation between
dust-to-gas ratio and metallicity among DLA systems (Vladilo 1998).
Given such a correlation, it is natural to expect that DLA points,
which have lower metallicities than interstellar clouds, have also
lower Cr depletion. The lack of any data points above the envelope
can be readily explained because clouds with low dust content
(and low Cr depletion) are expected to have low molecular content.
It is interesting that the data of the Magellanic Clouds fall
in the range occupied by DLAs, as expected. Indeed, typical 
dust-to-gas ratios in the Magellanic Clouds are
about 4 times (LMC) and 8 times (SMC) lower as compared with the
Milky Way (Koornneef 1982; Bouchet et al. 1985, respectively).

The fact that the DLA measurements lie on the envelope while
the upper limits lie below is puzzling. One possible explanation
is the presence of some selection effect that allows us to
detect H$_2$ at high Cr depletions only when the H$_2$ lines
are sufficiently strong. This could be the case if dust
obscuration is important.

The dust number density, $n_{\rm d}$, in the $z_{\rm abs} = 3.3901$
cloud may be estimated from the measured H$_2$ fractional abundance
assuming that the H$_2$ formation rate on dust grains, $R$, is
in equilibrium with the H$_2$ destruction caused by UV photons.
Another channel of the H$_2$ formation through the formation
of the catalyst H$^{-}$ (e.g., Donahue \& Shull 1991) may compete
with the dust process in the partially ionized warm regions and
will not be considered here because of the very high \ion{H}{i}
column density in the $z_{\rm abs} = 3.3901$ cloud which provides
most of the gas to be neutral. In the mostly neutral gas, H$_2$
will be in equilibrium with a density (e.g., Spitzer 1978)~:

\begin{equation}
I\,n({\rm H}_2) = R\,n\,n({\rm H})\: ,
\label{eq:E6}
\end{equation} 
where $I$ is the H$_2$ photo-dissociation rate in s$^{-1}$, 
$n = n({\rm H}) + 2\,n({\rm H}_2)$ in cm$^{-3}$, and $R$ in
cm$^3$s$^{-1}$.
For tenuous interstellar clouds, $I \simeq 0.11\,\beta_0$, with 
$\beta_0$ being the photo-absorption rate in the Lyman- and Werner-
bands. 

The H$_2$ formation rate on dust grains can be written as (Spitzer 1978)~:
\begin{equation}
R\,n = \frac{1}{2}\,\langle \gamma v_{\rm H} \rangle\, 
\langle n_{\rm d} \sigma_{\rm d} \rangle \: ,
\label{eq:E7}
\end{equation} 
where the constant $\gamma \simeq 0.3$, the mean thermal velocity of the
\ion{H}{i} atoms caused by particle collisions
$\langle v_{\rm H} \rangle \simeq 1.46\times10^4\,\sqrt{T_{\rm kin}}$\,
in cm~s$^{-1}$, and
$\langle n_{\rm d} \sigma_{\rm d} \rangle$ is the mean area obscured
by dust grains in 1~cm$^3$ along the sightline.

The value of $\beta_0$ depends on the local UV radiation field.
The intergalactic background radiation at $z \sim 3$ gives
$\beta_0 = 2\times10^{-12}$ s$^{-1}$ if $J_{21}$(912\,\AA) = 1
(Srianand \& Petitjean 1998). If we take into account that
locally in the intervening galaxy the UV emission from hot stars
may increase $\beta_0$ by a factor of $\kappa$, then using
(\ref{eq:E6}) and (\ref{eq:E7}) we have an estimate
\begin{equation}
\langle n_{\rm d} \sigma_{\rm d} \rangle =
5.0\times10^{-17}\frac{\kappa}{\sqrt{T_{\rm kin}}}\,
f({\rm H}_2)\: ,
\label{eq:E8}
\end{equation} 
where we took $n({\rm H}_2) / n({\rm H}) = N({\rm H}_2) / N(\ion{H}{i}$)
which is valid for the Q0000--2620 cloud.

Given the mean Galactic value 
$\langle n_{\rm d} \sigma_{\rm d} \rangle_{\rm ISM} = 1.2\times10^{-21}$
cm$^{-1}$\, (Spitzer 1978),\, one has
\begin{equation}
\frac{\langle n_{\rm d} \sigma_{\rm d} \rangle_{\rm DLA}}
{\langle n_{\rm d} \sigma_{\rm d} \rangle_{\rm ISM}} =
4.0\times10^4\frac{\kappa}{\sqrt{T_{\rm kin}}}\,
f({\rm H}_2)\: .
\label{eq:E9}
\end{equation} 

\begin{figure}
\hspace{0.0cm}\psfig{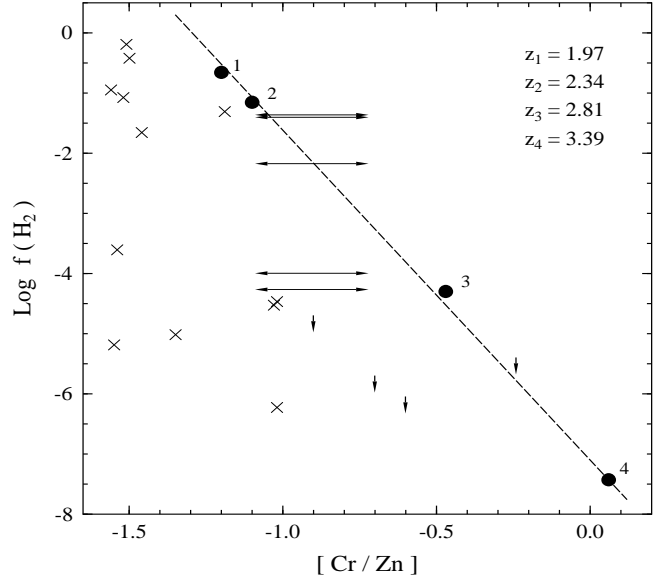}
\vspace{0.0cm}
\caption[]{Relation between H$_2$ fractional 
abundance plotted on a logarithmic scale and relative heavy element
depletion in high redshift DLAs (dots, vertical arrows label 
systems from Table~1).
The dashed line corresponds to the linear law~:
$\log f({\rm H}_2) \simeq -5.5\, [{\rm Cr}/{\rm Zn}] - 7.1$ .
The same plot for the Galactic ISM diffuse
clouds based on the data from the compilation of Roth \& Blades (1995)
are shown by crosses. 
The horizontal lines mark the H$_2$ measurements (Richter 2000) in the
Magellanic Clouds (see text for details).
References: 1,2~-- Ge \& Bechtold 1999;
3~-- Levshakov \& Varshalovich 1985, Srianand \& Petitjean 1998;
4~-- this work  
}
\end{figure}

If\, $\kappa / \sqrt{T_{\rm kin}} \sim 1$\, and the linear size
distribution function for the dust grains is invariant, then
for the $z_{\rm abs} = 3.3901$ cloud the last relation yields
$\langle n_{\rm d} \rangle_{\rm DLA} / \langle n_{\rm d} \rangle_{\rm ISM} \sim
10^{-3}$. 

It is reasonable to compare this quantity with
the dust-to-gas ratio, $\tilde{k}$, which
may be estimated from any two elements X and Y
having similar nucleosynthetic history (cf. Vladilo 1998)~:
\begin{equation}
\tilde{k} = \frac{10^{[{\rm X}/{\rm H}]_{\rm obs}}}
{f_{\rm X, ISM} - f_{\rm Y, ISM}}\left(
10^{[{\rm Y}/{\rm X}]_{\rm obs}} - 1\right)\: ,
\label{eq:E9a}
\end{equation} 
where the fraction in dust $f_{\rm X, ISM}$ refers to the
Galactic interstellar medium.

By applying error propagation to the column density measurements
for the \ion{Fe}{ii}$\lambda 1611$, \ion{Cr}{ii}$\lambda 2056$, and
\ion{Zn}{ii}$\lambda 2026$ lines
(Table~1 in Paper~I), one can find that [Fe/Zn] $= 0.03 \pm 0.07$
and [Cr/Zn] $= 0.05 \pm 0.08$. 
Equation (\ref{eq:E9a}) implies that positive values of the [Fe/Zn]
and [Cr/Zn] ratios are unphysical because of the dust depletions
$f_{\rm Zn, ISM} = 0.35$, $f_{\rm Cr, ISM} = 0.92$, and
$f_{\rm Fe, ISM} = 0.94$. 
Therefore we can estimate only maximum value of $\tilde{k}$
taking low bounds of [Fe/Zn] = $-0.04$ and [Cr/Zn] $= - 0.03$,
and the upper bound of [Zn/H] $= -1.97$. 
Under these conditions
we have $\tilde{k} \leq 1.6\times10^{-3}$. 
This value corresponds precisely to the dust number density ratio 
estimated above from the H$_2$ fractional abundance.
The same result implies that the current interpretation of the
Cr/Zn ratios as due to differential dust depletion rather than
intrinsic chemical abundance differences of Zn, as claimed
by several authors, is correct.

\section{ Future Prospects }
 
The accuracy of the estimated physical parameters $N({\rm H}_2)$ and
$T_{\rm ex}$ may be improved by additional observations of the quasar
0000--2620 to increase the signal-to-noise ratio at the expected
positions of the L(1-0)R(2) and L(9-0)R(4) lines where we observe
rather narrow ($| \Delta v | \la 15$ km~s$^{-1}$) continuum windows
free from the Ly$\alpha$ forest contaminations (see Fig.~4). At some
level of S/N one may detect weak absorption features corresponding to
these H$_2$ transitions. For the measured ortho-H$_2$ column density
$N_1 = 3.9\times10^{13}$ cm$^{-2}$, such detection at $3\sigma$
level becomes possible starting from S/N $\simeq 60$, if 
$T_{\rm ex} \ga 10^3$~K (see Table~2). The S/N ratios listed in 
Table~2 were estimated in the following way.

For weak absorption lines ($\tau_0 \ll 1$) one has (see LCFB)~: 
\begin{equation}
W \simeq 8.85\times10^{-18}\,N\,\lambda^2_0\,f_{\rm abs}\: ,
\label{eq:E10}
\end{equation} 
where the equivalent width $W$ is given in m\AA,\, $N$ in cm$^{-2}$,
and $\lambda_0$ in \AA.

The equivalent width detection limit, $\sigma_{\rm lim}$, defined in
LCFB, depends on the spectral resolution, the mean S/N value per pixel
and the accuracy of the continuum fit, $\sigma_{\rm c} / C$, over the
width of the line~: 
\begin{equation}
\sigma_{\rm lim} = \Delta \lambda\, \left[\,
\left(\frac{S}{N}\right)^{-2}\,{\cal M} +
\left(\frac{\sigma_{\rm c}}{C}\right)^2\,{\cal M}^2\,
\right]^{1/2}\: ,
\label{eq:E11}
\end{equation} 
where ${\cal M} = 2.5\times$FWHM is the full width (in pixels) at the
continuum level of a weak absorption line, and $\Delta \lambda$ is the
pixel size in \AA.

For real data, $\sigma_{\rm c}/C < ({\rm S}/{\rm N})^{-1}$
since the continuum level is usually estimated as a mean intensity
over $m$ pixels within a continuum window. Thus we may write
$\sigma_{\rm c}/C = \xi({\rm S}/{\rm N})^{-1}$, where 
$\xi \simeq 1/\sqrt{m}$.\, Then from (\ref{eq:E11}) we have
\begin{equation}
\frac{{\rm S}}{{\rm N}} = \frac{\Delta \lambda}{\sigma_{\rm lim}}\, 
\left[\,{\cal M}(1 + \xi^2 {\cal M}^2\,\right]^{1/2}\: .
\label{eq:E12}
\end{equation} 

In our calculations we set $\Delta \lambda = 0.04$ \AA,\,
FWHM = 2.5 pixels, and $\xi = 0.2$.
After that we calculate $N_{J=2}$ and $N_{J=4}$ using (\ref{eq:E5})
for a given $T_{\rm ex}$ and the corresponding $W$ values from
(\ref{eq:E10}). Taking $\sigma_{\rm lim} = \frac{1}{3}\,W$, we  
find S/N values from (\ref{eq:E12}).
The obtained results listed in the sixth and seventh columns of Table~2
show that the L(1-0)R(2) and/or L(9-0)R(4) lines may be detected for
a reasonable exposure time if $T_{\rm ex}$ is not very low in
the $z_{\rm abs} = 3.3901$ absorbing cloud.

Another issue we would like to outline in this section is the DLAs
with $N(\ion{H}{i}) \geq 10^{21}$ cm$^{-2}$ and $z_{\rm abs} > 2$
from the compilation of Ge \& Bechtold (1999). These systems are
listed in Table~1. Similar to our case, they show high neutral
hydrogen column densities but no H$_2$ absorption in moderate
resolution spectra (FWHM $\simeq 1$ \AA). The upper limits for
$f({\rm H}_2)$ are shown in the fifth column of Table~1.
In the sixth column we present the expected $f({\rm H}_2)$ values 
based on the linear relationship between $\log f({\rm H}_2)$
and [Cr/Zn] as found in this paper.
As noted in Sec.~2.3,
it is surprising that there is only one DLA system ($z_{\rm abs} = 2.30$)
which is consistent with our results, i.e. the predicted H$_2$ fractional
abundance is lower than the previously obtained upper limit. All other
systems show too low limits for $f({\rm H}_2)$ to be in concordance with
expected values.

\begin{table}
\centering
\caption{QSO candidates for H$_2$ absorption} 
\label{tab1}
\begin{tabular}{cccccc}
\hline
\noalign{\smallskip}
QSO & $z_{\rm abs}$ & $N(\ion{H}{i})$, & [Cr/Zn] &
$f^{a}({\rm H}_2)$ & $f^{b}({\rm H}_2)$  \\
\noalign{\smallskip}
 & & cm$^{-2}$ & & & \\ 
\hline
\noalign{\smallskip}
0458--02 & 2.04 & 8.0(21) & $-0.7$ & $<2.0(-6)$ & $5.0(-4)$ \\
0100+13  & 2.30 & 2.5(21) & $-0.24$& $<4.0(-6)$ & $1.6(-6)$ \\
0112+03  & 2.42 & 1.0(21) & $-0.9$ & $<2.0(-5)$ & $6.3(-3)$ \\
1223+17  & 2.47 & 3.0(21) & $-0.6$ & $<9.0(-7)$ & $1.6(-4)$ \\ 
\noalign{\smallskip}
\hline
\noalign{\smallskip}
\multicolumn{6}{l}{$^a$Upper limits from Ge \& Bechtold (1999)}\\
\multicolumn{6}{l}{$^b$Estimated from 
$\log f({\rm H}_2) = -5.5\,[{\rm Cr}/{\rm Zn}] - 7.1$}
\end{tabular}
\end{table}
\begin{table}
\centering
\caption{Predicted equivalent widths for the L(1-0)R(2)
and L(9-0)R(4) lines for different $T_{\rm ex}$ and the 
S/N ratios which can provide $3\sigma$ level detection
}
\label{tab2}
\begin{tabular}{ccccccc}
\hline
\noalign{\smallskip}
$T_{\rm ex}$ & $N_{J=2}$ & $N_{J=4}$ & $W^{\rm obs}_{{\rm R}(2)}$ &
$W^{\rm obs}_{{\rm R}(4)}$ & S/N & S/N \\
\noalign{\smallskip}
(K) & (cm$^{-2}$) & (cm$^{-2}$) & (m\AA) & (m\AA) & R(2) &
R(4) \\
\noalign{\smallskip} 
\hline
\noalign{\smallskip}
1400 & 1.70(13) & 1.30(13) & 2.90 & 7.97 & 165 & 60 \\
1000 & 1.54(13) & 8.39(12) & 2.63 & 5.14 & 180 & 90 \\
500  & 1.09(13) & 1.80(12) & 1.86 & 1.10 & 260 & 430 \\
\noalign{\smallskip}
\hline
\end{tabular}
\end{table}

One may advocate an intrinsic spread of the H$_2$ fractional abundance
among DLAs at constant [Cr/Zn] ratios as an explanation. It would 
be also necessary to investigate the sample of QSOs of Table~1  with data of 
similar resolution and S/N ratios than the one presented here.
Such studies could help us to
understand the dust formation history and the metal enrichment of
damped QSO systems since dust is mainly produced in the cool envelopes
of intermediate-mass stars and in the dense shells of supernova remnants.

\section{ Conclusions }

The outcome of our study, aiming to provide a measurement
of the molecular hydrogen abundance 
in the low metallicity DLA
galaxy at $z_{\rm abs} = 3.3901$ toward Q0000-2620, 
is as follows.

1. In the high resolution UVES spectrum of the quasar 0000--2620 
a weak absorption feature at $\lambda_{\rm obs} = 4609.4$ \AA\,
is identified with the H$_2$ L(4-0)R(1) line from the 
$z_{\rm abs} = 3.3901$ system.

2. The application of the standard Voigt fitting analysis
to this line gives the ortho-H$_2$ column density of
$N_{J=1} = (3.9 \pm 0.9)\times10^{13}$ cm$^{-2}$ 
and the Doppler width
$b_{{\rm H}_2} = 8.8^{+3.0}_{-0.8}$ km~s$^{-1}$.
The width of the H$_2$ line is found to match those 
for numerous unsaturated metal
lines measured at the same redshift in Paper~I. 

3. The upper limits to the column densities of the
W(0-0)R(0)+R(1), L(1-0)R(2), L(10-0)P(3), L(9-0)R(4), 
L(9-0)R(5) lines together with the measured value of $N_{J=1}$
lead to the excitation temperature $T_{\rm ex}$ ranging
from 50~K to 1400~K.

4. The range of $T_{\rm ex}$ and the $N_{J=1}$ value yield, in turn,
the total H$_2$ column density of
$N({\rm H}_2) \simeq (0.5 - 1.7)\times10^{14}$ cm$^{-2}$
and the H$_2$ fractional abundance of
$f({\rm H}_2) \simeq 4\times10^{-8}$.
This is the lowest H$_2$ fractional abundance ever measured.

5. Assuming that the H$_2$ formation rate on dust grains
is in equilibrium with the H$_2$ destruction caused by UV
photons, the dust number density is estimated
in the $z_{\rm abs} = 3.3901$ cloud, 
$\langle n_{\rm d} \rangle_{\rm DLA} \sim 10^{-3}\,
\langle n_{\rm d} \rangle_{\rm ISM}$.
This value is in excellent agreement with the dust-to-gas
ratio of $\tilde{k} \leq 1.6\times10^{-3}$ found independently 
from the [Cr/Zn] and [Fe/Zn] abundance ratios.
 
6. The derived H$_2$ fractional abundance excellently fits in the
linear relation between $\log f({\rm H}_2)$ and 
[Cr/Zn] measured for other DLAs at high redshift.

7. It is shown that future observations with higher signal-to-noise
ratio (S/N $ > 60$) may reveal a few additional H$_2$ lines from
the $z_{\rm abs} = 3.3901$ system which may increase the accuracy
of the $N({\rm H}_2)$ and $T_{\rm ex}$ measurements.

\begin{acknowledgements}
We are deeply indebted to the people involved in the realization 
of the UVES spectrograph which made possible to achieve the
results presented here.
We also thank an anonymous referee for his/her comments.
The work of S.A.L. is supported by the RFBR grant No.~00-02-16007.
\end{acknowledgements}

\end{document}